\documentclass[prl,preprint,showkeys,showpacs,preprintnumbers,amssymb,amsmath,superscriptaddress]{revtex4}
\usepackage{graphicx}

\begin{document}

\title{Ab initio molecular dynamics calculations of threshold displacement energies in silicon carbide}

\date{\today}

\author{G. Lucas}

\email{guillaume.lucas@etu.univ-poitiers.fr}

\author{L. Pizzagalli}

\affiliation{Laboratoire de M\'etallurgie Physique, CNRS UMR 6630, Universit\'e
de Poitiers,  B.P. 30179, 86962 Futuroscope Chasseneuil Cedex, France}

\begin{abstract}

Using first principles molecular dynamics simulations, we have determined the
threshold displacement energies and the associated created defects in cubic
silicon carbide. Contrary to previous studies using classical molecular
dynamics, we found values close to the experimental consensus, and also created
defects in good agreement with recent works on interstitials stability in
silicon carbide. We carefully investigated the limits of this approach. Our work
shows that it is possible to calculate displacement energies with first
principles accuracy in silicon carbide, and suggests that it may be also the case for other covalent materials. 

\end{abstract}



\preprint{report number}

\maketitle

Particle irradiation is a well known and extensively used technique, allowing to modify mechanical, magnetic, electrical and optical properties of materials. For instance, a suitable ion irradiation may harden a material, lead to a local oxydation state, or activate a magnetic order. The utility of ion irradiation is also well known for electronics, with the doping or gettering processes, and for radiation therapy. Besides, damage accumulation due to irradiation is also an important research field, related to space and nuclear applications. 

The interaction of an energetic ion with the matter is a complex phenomenon, especially at high energies. Impinging ions are simultaneously slowed down by inelastic collisions with electrons, and by elastic collisions with atoms. The displacement of lattice atoms leads to creation of defects and accumulation of damage. A key quantity, relevant to the process and different for each irradiated material, is the threshold displacement energy ($E_d$). $E_d$ may be defined as  the minimal kinetic energy that has to be transferred to a lattice atom in order to create a stable Frenkel pair that survives at least $10^{-12}$~s. For instance, $E_d$ values are required as key input in large-scale irradiation simulation packages, such as SRIM/TRIM, extensively used for determining implantation profiles in doping processes, or for calculating damage accumulation in materials. 

This quantity is rather difficult to measure, since single created defects have to be identified during experiments, and associated with a well-defined irradiation energy. Then, there has been an increasing number of works aiming at the $E_d$ determination from molecular dynamics simulations. The procedure is simple: after a defined impulse given to an atom, which is usually called the primary knock-on atom (PKA), the evolution of the system is monitored. Once the transfered energy exceeds the $E_d$, there is formation of a Frenkel pair in the system. As far as we know, all simulations but one \cite{Win98NIMB} have been done with molecular dynamics and classical empirical potentials. In fact, several reasons hinder ab initio molecular dynamics. First, determining an energy threshold from molecular dynamics requires many runs, since the kinetic impulsion is progressively increased to find the threshold, and the procedure is stochastic due to non zero temperature. Second, large systems must be employed for high $E_d$. Finally, there is usually a $E_d$ associated with each cristallographic direction and with each element in a multicomponent system, which considerably increases the number of runs. 

The relevant question is: are the empirical potentials precise enough to allow a correct determination of $E_d$? For metals, it seems that this is the case, embedded-atom potentials being able to model metallic bonds with a good accuracy \cite{Pas02PMA,Zep03PRB}. There is less certainty for other materials, such as semiconductors or ceramics, since covalent bonds or charge transfers are hard to reproduce with potentials. In that case, an ab initio molecular dynamics 
determination would be extremelely useful. 

The silicon carbide is a good illustration of this issue. It is a promising material, with potential applications in electronics, as a replacement for silicon,  and in nuclear technology. Silicon carbide is also very interesting from a fundamental point of view, since it can be considered as a model for zinc-blende two-component covalent materials. There have been several measurements of the $E_d$, with different techniques, but a large dispersion of values is obtained \cite{Zin97JNM}. In lack of precise data, it is usually assumed that average values for C and Si sublattices are 20~eV and 35~eV, respectively. However, subsequent molecular dynamics studies did not clearly confirm these values. Average values were found from 17 to 40~eV for C sublattice and from 42 to 57~eV for Si sublattice, with extreme values very different \cite{Win98NIMB,Per97JNM,Dev98JNM,Dev00JNM,Per00JNM,Mal02PRB}. In addition, the nature of the created defects is different from one study to another. We have recently shown that these discrepancies are due to the use of different empirical potentials \cite{Luc05NIMB}. In fact, the kinetic energy required for the creation of a Frenkel pair is obviously related to the energy barrier that the lattice atom must overcome to reach an interstitial site. Empirical potentials usually give a poor description of these saddle states, especially for covalent materials. A precise ab initio determination would be invaluable in that case. 

\begin{figure}
\includegraphics[width=10cm]{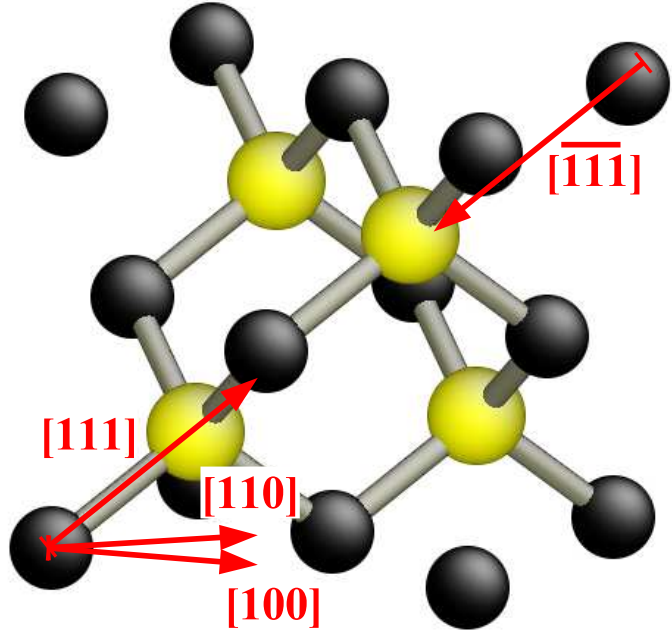}
\caption{Representation of the main crystallographic directions in $\beta$-SiC. Carbon atoms are drawn in black, and silicon atoms in light grey (yellow in the electronic version).}
\label{3C-SiC_cryst_dir}  
\end{figure}

In this paper, we report the first ab initio molecular dynamics determination of $E_d$. On the one hand, we show that such calculations are feasible, at least for covalent materials for which the vacancy-interstitial separation of the Frenkel pair is very small. On the other hand, $E_d$ values have been obtained in $\beta$-SiC for all high symmetry directions shown in figure \ref{3C-SiC_cryst_dir}, for both Si and C lattices, with the first principles accuracy. Our results show that the use of available empirical potentials may lead to quantitative and qualitative errors, and that our calculated average values are close to the experimental consensus. 

The ab initio molecular dynamics calculations were performed using the plane-wave pseudopotential code GP \cite{JEEP}, based on the density functional theory (DFT) \cite{Hoh64PRB,Koh65PRA}. The exchange-correlation potential proposed by Ceperley and Alder, and parametrized by Perdew and Zunger was used \cite{Per81PRB}. We considered a  $\Gamma$-sampling of the Brillouin zone, and a 35~Ry kinetic cut-off. With those parameters, the calcultated lattice parameter $a_0=4.34$ \AA~and the bulk modulus $B=221$ GPa were found to reproduce rather well experimental values, 4.36~\AA~and 224~GPa respectively \cite{LB17C}. We also checked that pseudopotential cores did not overlap during simulations. All calculations were performed with a constant number of particules, with a 64-atom cell ($2a_0\times2a_0\times2a_0$), except for the Si PKA in the $\langle100\rangle$ direction  where a 96-atom cell ($3a_0\times2a_0\times2a_0$) was required to keep the PKA in the cell. A time step $dt=1~a.u$ was used during the ballistic phase of the simulation, then increased to $2~a.u$ during the relaxation phase. A thermostat was applied to recover the initial temperature of 300~K during the latter phase.  The maximum duration of each run was 2.8~ps. If a stable Frenkel pair occured, the system was then completely relaxed to obtain the stable configuration.

\begin{figure}
\includegraphics[width=16cm]{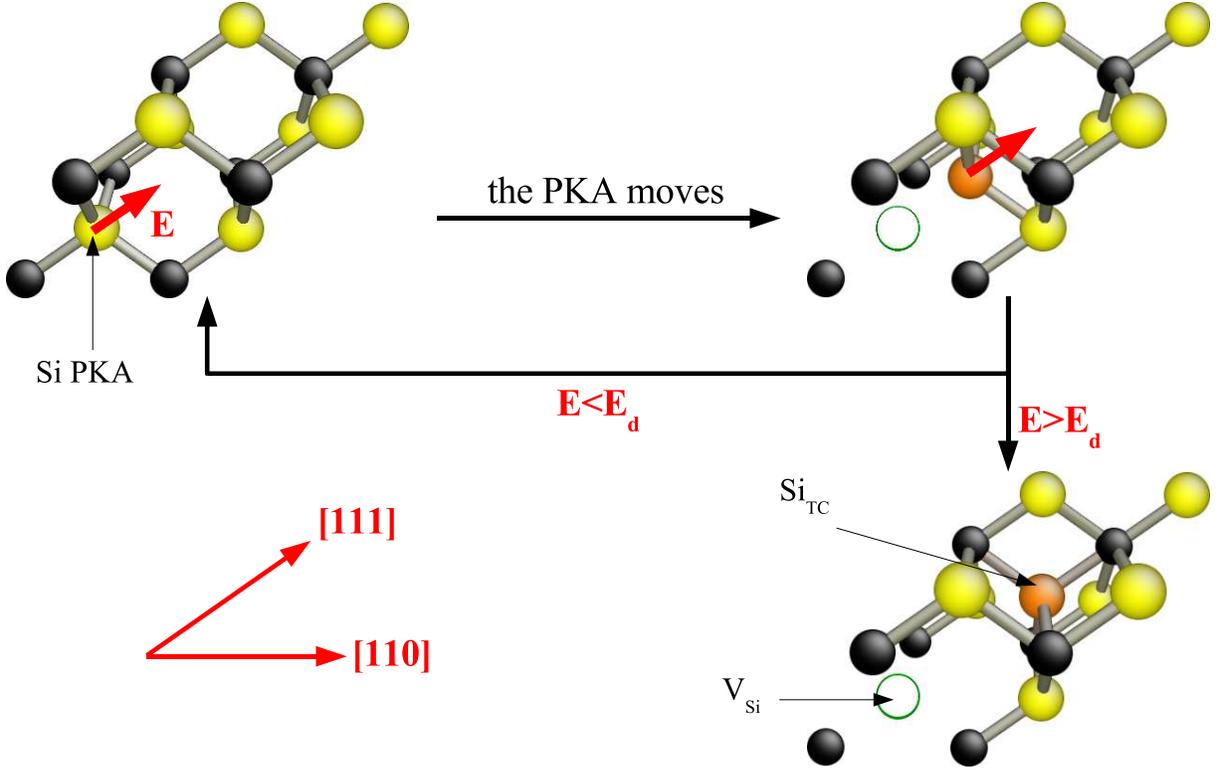}
\caption{A Si PKA along the $\langle111\rangle$ direction. Carbon atoms are drawn in black, and silicon atoms in light grey (yellow in the electronic version). The silicon PKA is drawn in grey (orange in the electronic version), and the vacancy is represented by an open circle. A kinetic energy $E$ is given to a Si atom,  which is subsequently displaced. If E\textless $E_d$, the PKA returns to its original location. If E\textgreater $E_d$, there is formation of a silicon vacancy $V_{Si}$ and a silicon tetrahedral interstitial surrounded by four carbon atoms $Si_{TC}$.}
\label{FP}  
\end{figure}

As an example, the figure \ref{FP} shows two possible cases in a typical threshold displacement energy determination, after a kinetic energy $E$ is transferred to a silicon atom along the $\langle111\rangle$ direction. The PKA first moves from its equilibrium position along the $\langle111\rangle$ direction. If E is below the threshold displacement energy $E_d$, in that case 22~eV, it returns to this location and no Frenkel pair is created. On the contrary, if $E$ is above $E_d$, the PKA reaches an interstitial location in the lattice, leaving its original site free. Thus there will be formation of a Frenkel pair, i.e an interstitial and a vacancy, separated by a distance $d_{FP}$. In this example, a vacancy and a silicon in a carbon tetrahedral site ($V_{Si}$+$Si_{TC}$), separated by a distance $d_{FP}=0.87a_0$, are produced above the $E_d$.

There are several computational issues that are supposed to prevent the determination of $E_d$ with first principles methods. Hence, the cell must be big enough to contain the PKA during all the simulation. Here, we have mainly used a 64-atom cell, which may be viewed as very small. However, in our simulations, the PKA does not move far away from its initial location before to be trapped in an interstitial site. Indeed in covalent materials, and especially in ceramics, the vacancy-interstitial separation $d_{FP}$ is very short, often lower than $a_0$. This is clearly in contrast to metals, for which $d_{FP}$ is several times $a_0$. Also, the cell should be large enough to prevent cumbersome interactions between the PKA and the thermostat during the simulation. Hence, in silicon, it has been suggested that a 64-atom cell is too small with respect to this issue \cite{Maz01PRB}. However we have recently shown that, in silicon carbide, the error due to the cell size problem is small compared to the discrepancy found between different calculation methods \cite{Luc05NIMB}. This is an important point, and we have performed an additional test with a larger cell (216-atom) and C$\langle100\rangle$ to check the validity of this assumption. We found no difference with the 64-atom cell, with a similar $E_d$ value. Another issue is related to the time step. It must be small enough to insure the accuracy of atomic trajectories, especially during the ballistic phase of the simulation.  Hence, we have used a time step of $1~a.u$, so that the maximum displacement during one time step for a C PKA of 50~eV is less than 0.007~\AA, which is much lower than the upper threshold of 0.1~\AA~recommended by Corrales \textit{et al.} for low energy cascade events \cite{Cor05NIMB}. Regarding all these points, we assert that the determination of $E_d$ by ab initio methods is feasible at least in ceramics, and, as it will be shown further, these calculations are required for determining accurately the threshold displacement energies and the created defects.

\begin{table}
\begin{tabular}{cccc}
\hline
Direction & $E_d$ (eV) & Defect & $d_{FP}$ ($a_0$)  \\
\hline
C$\lbrack100\rbrack$ & 18 & $V_C+tilted~CC_{\langle100\rangle}$ & 0.87 \\
C$\lbrack110\rbrack$ & 14 & $V_C+CSi_{\langle0\bar{1}0\rangle}$ & 0.48 \\
C$\lbrack111\rbrack$ & 38 & /  & /  \\
C$\lbrack\bar{1}\bar{1}\bar{1}\rbrack$ & 16 & $V_C+CSi_{\langle010\rangle}$ & 0.95 \\
\hline
\multicolumn{4}{c}{C sublattice, weighted average: 19~eV} \\
\hline
Si$\lbrack100\rbrack$ & 46 & $V_{Si}+Si_{TC}$ & 1.52 \\
Si$\lbrack110\rbrack$ & 45 & $V_C+CSi_{\langle0\bar{1}0\rangle}$ & 0.48 \\
Si$\lbrack111\rbrack$ & 22 & $V_{Si}+Si_{TC}$ & 0.87  \\
Si$\lbrack\bar{1}\bar{1}\bar{1}\rbrack$ & 21 & $V_C+CSi_{\langle0\bar{1}0\rangle}$ & 1.24 \\
\hline
\multicolumn{4}{c}{Si sublattice, weighted average: 38~eV} \\
\hline
\end{tabular}
\caption{Threshold displacement energies in $\beta$-SiC, calculated by DFT-LDA molecular dynamics, along the main crystallographic directions. The associated defects and resulting Frenkel pair separations $d_{FP}$ are also added. $V_C$, $V_{Si}$, $CC$, $CSi$ and $Si_{TC}$ correspond respectively to a carbon vacancy, a silicon vacancy, a carbon-carbon dumbbell, a carbon-silicon dumbbell and a silicon in a carbon tetrahedral site. The average values are weighted for equivalent directions. For the C$\lbrack111\rbrack$ case, several defects were observed.}
\label{TDE_DFT}
\end{table}

\begin{figure}
\includegraphics[width=18cm]{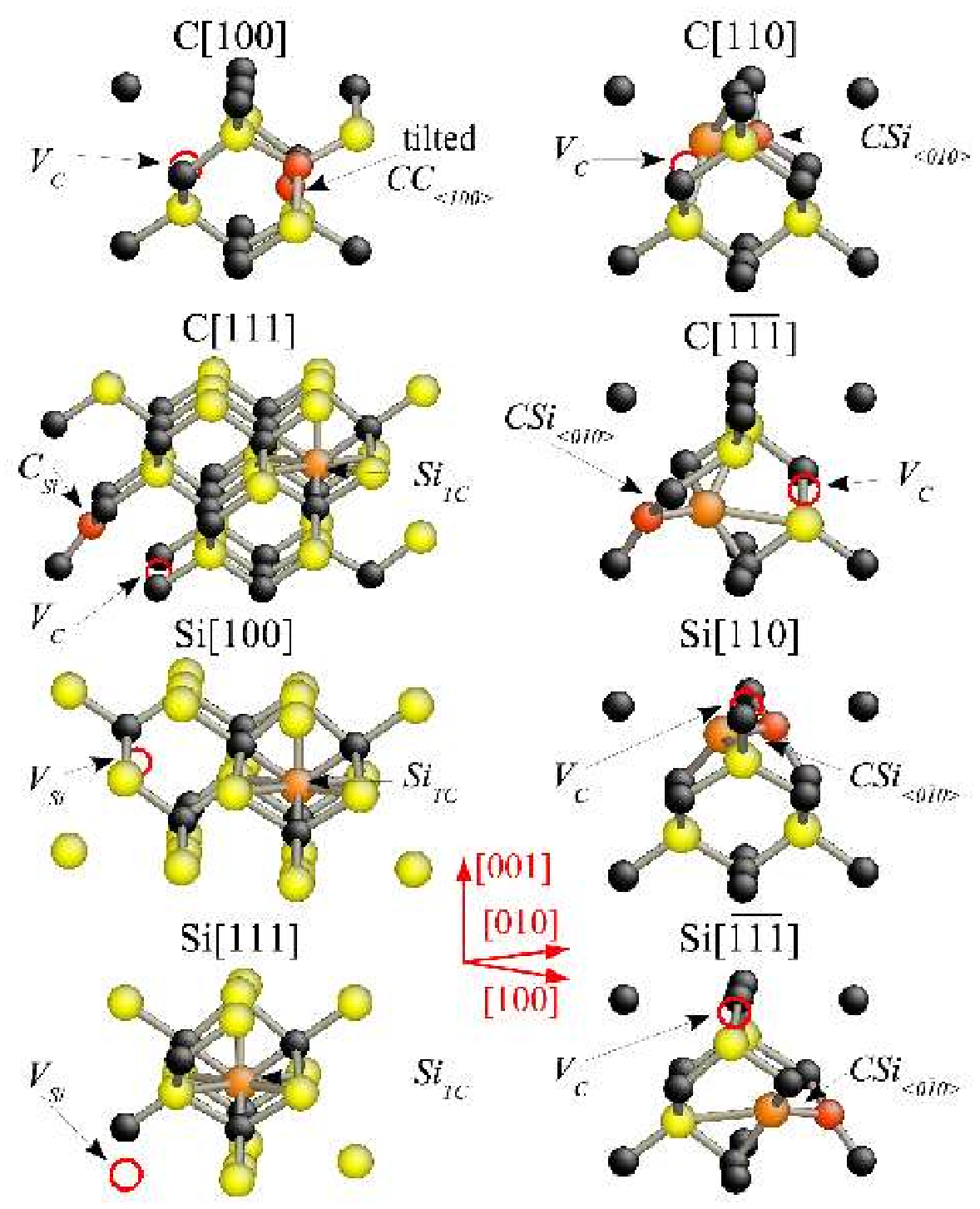}
\caption{Defect configurations for each considered crystallographic directions. Carbon atoms are drawn in black, and silicon atoms in light grey (yellow in the electronic version). Defects are drawn in grey (orange and red  for C and Si atoms, respectively, in the electronic version), and the vacancies are represented by an open circle.}
\label{defects}  
\end{figure}

We now describe and discuss our results. Table \ref{TDE_DFT} reports the calculated $E_d$ values and the associated Frenkel pairs, obtained for PKA's on both C and Si sublattices in the main cristallographic directions. The corresponding defect configurations are reproduced in figure \ref{defects}. Globally, our results show  various dumbbells and Si interstitials in tetrahedral site $Si_{TC}$. For C$\lbrack100\rbrack$ and an energy  above 18~eV, the PKA recoils toward the nearest tetrahedral interstitial site and moves further until it forms a tilted $CC_{\langle100\rangle}$ dumbbell interstitial with $d_{FP}$ equal to $0.87a_0$. This configuration was previously described as the most stable $CC$ dumbbell \cite{Boc03PRB}. Several $CSi$ dumbbells were also identified. For C$\lbrack110\rbrack$ and $Ed$ equal to 14~eV, the C atom replaces its C first neighbor, which is subsequently displaced to create a $CSi_{\langle0\bar{1}0\rangle}$ with $d_{FP}=0.47a_0$.  This configuration is also found in the case of a Si PKA along the $\langle110\rangle$ direction, and an energy above 45~eV, with a different collision sequence. Considering now C$\lbrack\bar{1}\bar{1}\bar{1}\rbrack$ direction, above 16~eV, the C atom heads for the tetrahedron defined by four Si atoms, and does not form a $C_{TSi}$ tetrahedral interstitial as it could be primarly expected, but a  slightly tilted $CSi_{\langle010\rangle}$ dumbbell with a Si atom. The Frenkel pair separation $d_{FP}$ is $0.95a_0$. This is consistent with previous ab initio calculations from Lento \textit{et al.}, predicting the conversion of the $C_{TSi}$ tetrahedral interstitial to the $CSi_{\langle010\rangle}$ dumbbell interstitial \cite{Len04JPCM}. The last case for which a $CSi$ dumbbell is obtained is the Si$\lbrack\bar{1}\bar{1}\bar{1}\rbrack$ with $E_d$ equal to 21~eV.  Here the Si atom collides with its C first neighbor, displaces it, and returns to its original location. The resulting $CSi_{\langle0\bar{1}0\rangle}$ interstitial is separated from the vacancy by $1.24a_0$. Silicon tetrahedral interstitials surrounded by carbon atoms $Si_{TC}$, which were determined as the most stable tetrahedral interstitial \cite{Boc03PRB,Len04JPCM,Sal04JNM}, were also created. The most simple case is Si$\lbrack111\rbrack$ described in figure \ref{FP}. Above 22~eV, the Si PKA directly moves toward the tetrahedral site and forms a $Si_{TC}$, $0.87a_0$ away from the vacancy. A Si PKA along the $\langle100\rangle$ direction, with an energy higher than 46~eV, leads to the formation of  a $Si_{TC}$ interstitial separated from the Si vacancy by $1.52a_0$, after a short collision sequence during which the Si PKA replaces another Si atom, this one moving in the following tetrahedral site. For the C$\lbrack111\rbrack$ case and an energy higher than 38~eV, several mechanisms, occuring for similar energies, were observed depending on the way the C PKA rebounded on its closest silicon neighbor. In the first mechanism, the C PKA rebounds without displacing the Si atom and forms $CSi_{\langle\bar{1}00\rangle}$, identical to C$\lbrack110\rbrack$  and  Si$\lbrack110\rbrack$ cases. In the others, the C PKA encounters its Si first neighbor at short distance with enough energy to displace it to the next $Si_{TC}$ interstitial site. Afterwards, the C PKA sometimes returns to its original location, leading to a final configuration similar to the Si$\lbrack111\rbrack$ case, or it bounces backward, and after few recombinations forms additional defects such as $C_{Si}$ antisite and carbon vacancy $V_C$, as shown in figure \ref{defects}. In this peculiar case, there is an uncertainty regarding the created defects, but for a similar $E_d$, a somewhat different result than in previous works \cite{Mal02PRB}. Finally, regarding all the different PKA's that have been studied, the created defects are always in fair agreement with the relative stability of defects found with static ab initio calculations.

We have determined the average $E_d$ on both C and Si sublattices, by weighting each values of $E_d$ by the number of equivalent directions \cite{Eqdir}. Our average $E_d$ are in very good agreement with the values usually considered by the fusion community: 19~eV against 20~eV for the C sublattice, and 38~eV against 35~eV for the Si sublattice.

Several previous investigations were devoted to the determination of threshold displacement energies in silicon carbide with classical molecular dynamics, but with differences in calculated $E_d$ values. Moreover, identified defects strongly diverged between all studies. Roughly it is possible to sort  previous results into two groups. The first one, related to the original Tersoff potential \cite{Per97JNM,Win98NIMB},  shows similar or slightly higher $E_d$ values than in our work, but the created Frenkel pairs, mostly $V_C+C_{TSi}$ with low vacancy-interstitial separations, seems unphysical. It could be explained by the fact that the original Tersoff potential highly favored the formation of the $C_{TSi}$ interstitial \cite{Sal04JNM,Hua95MSMSE}. Conversely, the second group, related to Tersoff potentials modified for short range interactions \cite{Dev98JNM,Dev00JNM,Per00JNM,Mal02PRB}, exhibits a more realistic defects production (essentially dumbbells), but also much higher $E_d$ than in our work. Thus it seems hard to achieve a good description of the defect production together with accurate values of $E_d$ using  semi-empirical potentials. Ab initio tight-binding molecular dynamics have already been performed for two crystallographic directions \cite{Win98NIMB}, but the results were not convincing as the authors used a minimal basis set unable to take into account charge transfers. As a result, calculated defect formation energies, as well as $E_d$ values, were not accurate. For example, the authors found $E_d$ equal to 27.5~eV for a C PKA along the $\langle100\rangle$ direction, against 18~eV, here. State of the art first principles calculations are then required for determining accurately both $E_d$ and formed defects.

In conclusion, we have demonstrated  that it is feasible to determine $E_d$ in silicon carbide using ab initio molecular dynamics. This method could also be applied for silicon and other covalent materials. Average calculated values of $E_d$ were found in very close agreement to the experimental consensus for both C and Si sublattices. Such an agreement, both with the experiment and the calculated defect formation energies, has never been found with semi-empirical potentials or tight-binding methods, and hence justifies the use of first principles methods for the determination of threshold displacement energies in covalent materials.


\begin{thebibliography}{22}
\bibitem[{Eqd()}]{Eqdir}
\bibinfo{note}{There are 6 equivalent $\langle100\rangle$ directions, 12
  $\langle110\rangle$, 4 $\langle111\rangle$, and 4
  $\langle\bar{1}\bar{1}\bar{1}\rangle$.}
\end{thebibliography}

\begin{acknowledgments}
This work was funded by the joint research program "ISMIR" between CEA and CNRS.
\end{acknowledgments}

\end{document}